\documentclass[aps,nofootinbib,groupaddress,floatfix,preprint]{revtex4-1}
\usepackage{graphicx}
\usepackage{amsmath}
\usepackage{dcolumn}
\usepackage{subfigure}
\usepackage{multirow}
\usepackage{mathtools}
\usepackage{epstopdf}
\usepackage{mathptmx}
\usepackage{braket,comment}
\begin{document}

\title{Approaching the Standard Quantum Limit of Mechanical Torque Sensing}

\author{P.H. Kim}\thanks{These authors contributed equally to this work.}
\author{B.D. Hauer}\thanks{These authors contributed equally to this work.}
\author{C. Doolin}
\author{F. Souris}
\author{J.P. Davis}\email{jdavis@ualberta.ca}
\affiliation{Department of Physics, University of Alberta, Edmonton, Alberta, Canada T6G 2E9}

%\affiliation{Department of Physics, University of Alberta, Edmonton, Alberta, Canada T6G 2E9}

%\date{\today}

\maketitle

\linespread{1.2}

\textbf{Mechanical transduction of torque has been key to probing a number of physical phenomena, such as gravity \cite{Cavendish1798}, the angular momentum of light \cite{Beth1936}, the Casimir effect \cite{Chan2001}, magnetism \cite{Davis2010,Rugar2004}, and quantum oscillations \cite{Lupien1999}.  Following similar trends as mass \cite{Chaste2012} and force \cite{Miao2012} sensing, mechanical torque sensitivity can be dramatically improved by scaling down the physical dimensions, and therefore moment of inertia, of a torsional spring \cite{Davis2010}. Yet now, through precision nanofabrication and sub-wavelength cavity optomechanics \cite{Kim2013,Wu2014}, we have reached a point where geometric optimization can only provide marginal improvements to torque sensitivity.  Instead, nanoscale optomechanical measurements of torque are overwhelmingly hindered by thermal noise.  Here we present cryogenic measurements of a cavity-optomechanical torsional resonator cooled in a dilution refrigerator to a temperature of 25 mK, corresponding to an average phonon occupation of $\mathbf{\braket{n} = 35}$, that demonstrate a record-breaking torque sensitivity of 2.9 yNm/$\sqrt{\textrm{Hz}}$.  This a 270-fold improvement over previous optomechanical torque sensors \cite{Kim2013,Wu2014} and just over an order of magnitude from its standard quantum limit.  Furthermore, we demonstrate that mesoscopic test samples, such as micron-scale superconducting disks \cite{Geim1997}, can be integrated with our cryogenic optomechanical torque sensing platform, in contrast to other cryogenic optomechanical devices \cite{Teufel2011,Wollman2015,Meenehan2015,Riedinger2016}, opening the door for mechanical torque spectroscopy \cite{Losby2015} of intrinsically quantum systems.}

Cavity optomechanics \cite{Aspelmeyer2014} allows measurement of extremely small mechanical vibrations via effective path length changes of an optical resonator, as epitomized by the extraordinary detection of the strain resulting from transient gravitational waves at LIGO \cite{Abbott2016}. Harnessing cavity optomechanics has enabled measurements of displacement (the basis for force and torque sensors) of on-chip mechanical devices \cite{Eichenfield2009,Anetsberger2009} at levels unattainable by previous techniques.  Here we focus on cavity optomechanics as a platform for measuring torque applied to a torsional spring. 

\begin{figure*}[t]
%%%%%%%%%%%%%%%%%   F I G U R E  1   %%%%%%z%%%%%%%%%%%%
\centerline{\includegraphics[width=6.3in]{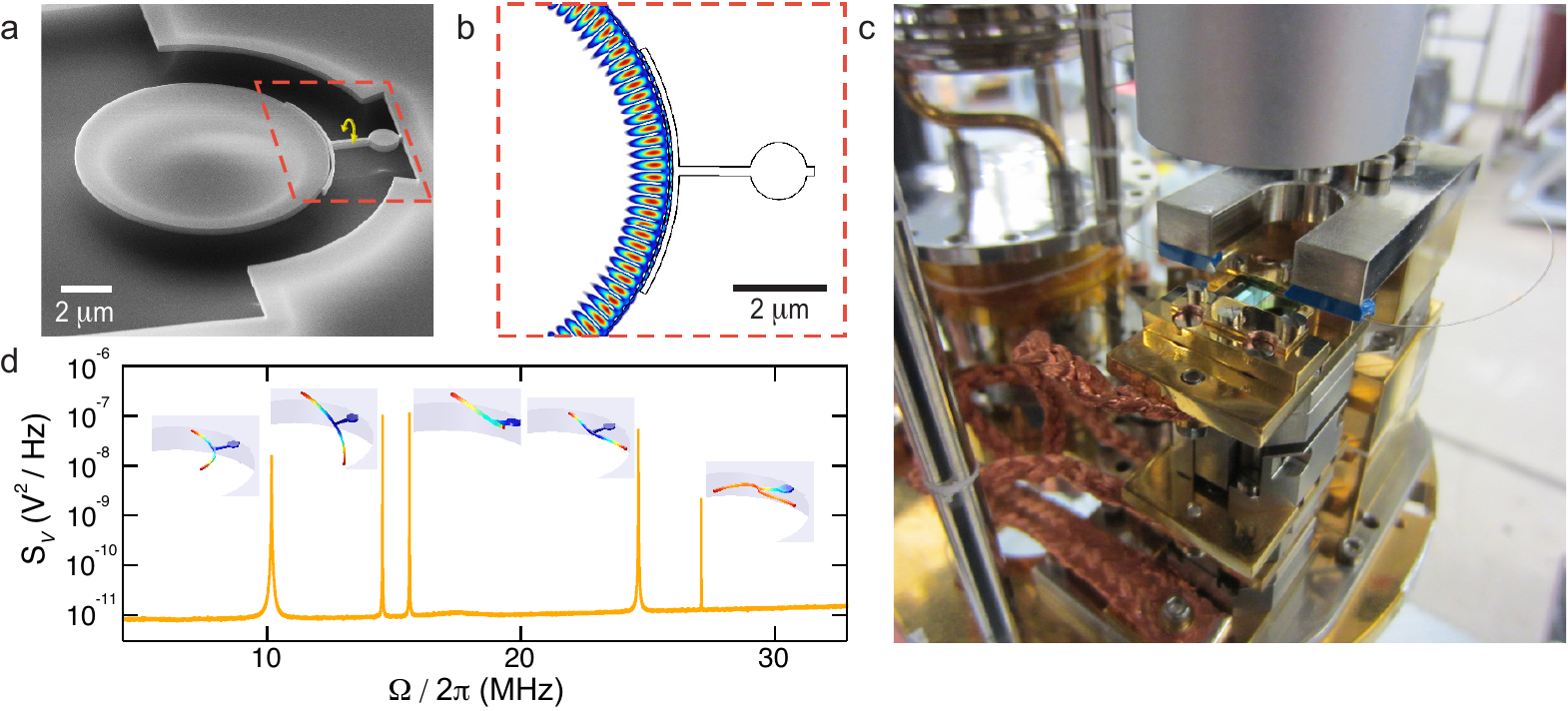}}
%%%%%%%%%%%%%%%%%%%%%%%%%%%%%%%%%%%%%%%%%%%%%
\caption{{\bf Low temperature optomechanics.} \linespread{1.0} \textbf{a}, Scanning electron micrograph of the optomechanical torque sensor used here: a 10 $\mu$m diameter optical microdisk evanescently coupled to a torsional nanomechanical resonator by a vacuum gap of 60 nm. A 1.1 $\mu$m diameter ``landing pad" allows for deposition of secondary test samples.  \textbf{b}, Finite element model of the microdisk optical resonance, which couples dispersively to the mechanical resonator. \textbf{c}, The sample chip is clamped into a gold-plated copper mount, on a stack of low-temperature-compatible nanopositioners, with thermal braids linking it to the base-plate of the dilution refrigerator \cite{MacDonald2015}. A dimpled-tapered optical fiber is held on a positionable Invar fork above the sample stage.  \textbf{d}, Optomechanically measured thermal noise voltage spectrum of the five lowest order mechanical modes, at 4.2 K, with finite element simulations of the mode shapes color coded by their total displacements.    {\label{fig1}} }
\end{figure*}

For a thermally-limited, classical measurement, the minimum resolvable mechanical torque spectrum is given by
\begin{equation}
S_{\tau}^\textrm{cl} = 4 k_{\textrm{B}} T \Gamma I,
\label{Staucl}
\end{equation}
where $k_{\textrm{B}}$ is the Boltzmann constant and $T$ is the mode temperature. By taking the square root of equation \eqref{Staucl}, one obtains the device's torque sensitivity (in units of Nm/$\sqrt{\rm Hz}$). Minimization of the (effective) moment of inertia, $I$, and the mechanical damping rate, $\Gamma$, can therefore result in improved torque sensitivity at a given temperature.  Reducing the mechanical damping is notoriously challenging, and even with modern nanofabrication techniques---paired with cavity optomechanical detection of sub-optical wavelength structures---the moment of inertia can only be lowered so far.  Here geometric optimization leads to the design shown in Fig.~\ref{fig1}a, with a room temperature torque sensitivity of 0.4 zNm/$\sqrt{\textrm{Hz}}$: a modest improvement over the 0.8 zNm/$\sqrt{\textrm{Hz}}$ of previous incarnations of cavity optomechanical torque sensors \cite{Kim2013,Wu2014}.  Therefore further improvement demands lowering of the mechanical mode temperature.

Fortunately, cavity optomechanics has been successfully integrated into cryogenic environments. Multiple architectures are now capable of cooling near the quantum ground state either directly through passive cooling \cite{OConnell2010,Meenehan2015,Riedinger2016}, or in combination with optomechanical back-action cooling \cite{Chan2011,Teufel2011,Wollman2015}.  Yet, as we  demonstrate below, only passive cooling is compatible with reducing the thermal noise of an optomechanical torque sensor.  Furthermore, while these architectures are well suited to tests of quantum mechanics and applications of quantum information processing, they are not well suited to integration with external systems one may wish to test.  Hence our cavity optomechanical torque sensing platform is unique, in that it enables straightforward integration with test samples \emph{and} operates in a dilution refrigerator with near quantum-limited torque sensitivity.

To understand the limit of torque sensitivity in the quantum regime, one must consider the device's intrinsic angular displacement spectrum, $S_{\theta}^{\rm qu}(\Omega)$, as well as the imprecision and back-action noise spectra associated with the measurement apparatus, $S_{\theta}^{\rm imp}(\Omega)$ and $S_{\theta}^{\rm ba}(\Omega)$, giving a total measured angular noise spectral density of
\begin{equation}
S_{\theta}(\Omega) = S_{\theta}^{\rm qu}(\Omega) + S_{\theta}^{\rm imp}(\Omega) + S_{\theta}^{\rm ba}(\Omega).
\label{Stot}
\end{equation}
The intrinsic noise spectrum can be expressed as $S_{\theta}^{\rm qu}(\Omega) = |\chi(\Omega)|^2 S_{\tau}^{\rm qu}(\Omega)$, where $\chi(\Omega)$ is the torsional susceptibility, and we introduce the quantum thermal torque spectrum, $S_{\tau}^{\rm qu}(\Omega) = 4 \hbar \Omega \Gamma I (\braket{n} + 1/2)$ (see Supplementary Information). This spectrum is comprised of both the average phonon occupation of the mechanical mode, $\braket{n}$, which for a resonator in equilibrium with an environmental bath is given by $\braket{n} = \bar{n}_{\rm th} = (e^{\hbar \Omega / k_B T} - 1)^{-1}$, as well as a ground state contribution, manifest as an addition of one half to the resonator's phonon occupation.

\begin{figure}[b]
%%%%%%%%%%%%%%%%%   F I G U R E  2   %%%%%%%%%%%%%%%%%%
\centerline{\includegraphics[width=3.5in]{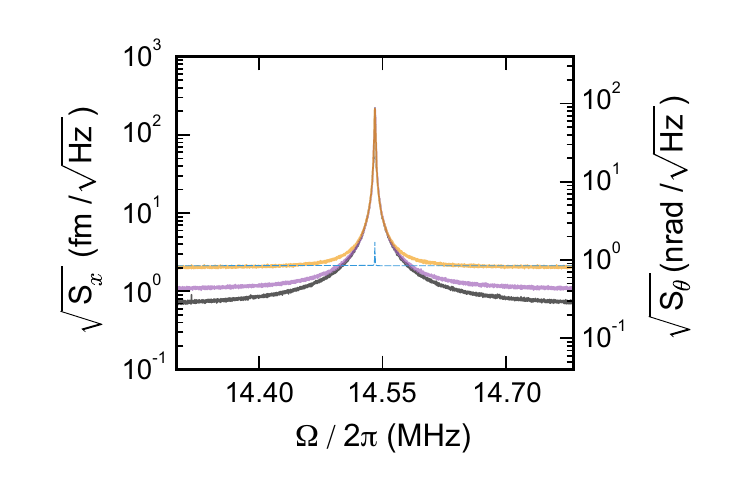}}
%%%%%%%%%%%%%%%%%%%%%%%%%%%%%%%%%%%%%%%%%%%%%
\caption{ {\bf Power dependence at 4.2 K.}  Optomechanically measured thermal noise spectra, calibrated in terms of linear and angular displacement, of the torsional mode in the presence of helium exchange gas.  As the optical power injected into the device is increased, the photon shot noise is reduced and the measurement imprecision drops below the noise floor of 0.84 nrad/$\sqrt{\rm Hz}$ (2.1 fm/$\sqrt{\rm Hz}$) corresponding to the standard quantum limit. Optical power input to the device (and average intracavity photon number) corresponds to: orange $33~\mu$W (6.0$\times 10^{3}$ photons), purple $86~\mu$W (1.3$\times 10^{4}$ photons), and grey $156~\mu$W (2.7$\times 10^{4}$ photons), resulting in angular (linear) displacement imprecision noise floors of 0.75 nrad/$\sqrt{\rm Hz}$ (1.9 fm/$\sqrt{\rm Hz}$), 0.39 nrad/$\sqrt{\rm Hz}$ (0.97 fm/$\sqrt{\rm Hz}$) and 0.25 nrad/$\sqrt{\rm Hz}$ (0.62 fm/$\sqrt{\rm Hz}$). The zero-point noise spectrum at the SQL, calculated using measured device parameters, is shown as the blue dashed line. {\label{fig2}} }
\end{figure}

Furthermore, the back-action and imprecision noise will result from a combination of both technical and fundamental noise associated with the measurement apparatus, whose product is bounded from below (for single-sided spectra \cite{Hauer2013}) by the Heisenberg uncertainty relation 
\begin{equation}
S_{\tau}^{\rm ba}(\Omega) S_{\theta}^{\rm imp}(\Omega) \ge \hbar^2,
\label{Heis}
\end{equation}
with equality corresponding to a measurement limited solely by quantum noise \cite{Clerk2010}. Using this relation, it is possible to determine the measurement strength at which the added fundamental back-action and imprecision noise will be minimized, corresponding to the standard quantum limit (SQL) of continuous linear measurement. Considering a torque at the resonance frequency, $\Omega_{\rm m}$, of the mechanical system---where the displacement signal is maximized---the minimum resolvable torque spectrum at the SQL is found to be
\begin{equation}
S^{\rm SQL}_{\tau} = \frac{S^{\rm SQL}_\theta(\Omega_{\rm m})}{|\chi(\Omega_{\rm m})|^2} = S_{\tau }^0 (\braket{n} + 1).
\label{Stauqu}
\end{equation}
Here, $S_{ \tau}^0 = 4 \hbar \Omega_{\rm m} \Gamma I$ is the fundamental torque noise limit associated with a continuous linear measurement of a mechanical resonator in its ground state. Half of this fundamental torque noise arises from the zero-point motion of the resonator, the other half from the Heisenberg-limited measurement noise at the SQL. Note that in the classical limit, $\braket{n} \approx k_{\rm B} T / \hbar \Omega_{\rm m} \gg 1$, both of these effects can be neglected and the minimized quantum torque noise of equation \eqref{Stauqu} is equivalent to its classical counterpart given by equation \eqref{Staucl}.

As illustrated by equation \eqref{Stauqu}, the torque sensitivity is improved as the average phonon occupancy of the device is decreased. Therefore, one might naively think to use some form of active cold damping to reduce the phonon occupancy of the mechanical resonator and consequently its minimum resolvable torque. For example, one could employ optomechanical back-action cooling (OBC), which has been successful in reducing the phonon occupancy of nanoscale mechanical resonators near to their quantum ground state \cite{Chan2011,Teufel2011,Wollman2015}. In OBC, a dynamical radiation pressure force imparted by photons confined to an optical cavity effectively damps the device's motion, increasing its intrinsic mechanical linewidth by an amount $\Gamma_{\rm OM}$ and reducing its average phonon occupancy according to
\begin{equation}
\braket{n} = \frac{\Gamma \bar{n}_{\rm th} + \Gamma_{\rm OM} \bar{n}_{\rm min}}{\Gamma + \Gamma_{\rm OM}},
\label{naveba}
\end{equation}
\noindent where $\bar{n}_{\rm min}$ is the minimum obtainable average phonon number using this method \cite{Aspelmeyer2014}. Inputting this expression for $\braket{n}$ into equation \eqref{Stauqu}, one can determine the minimum resolvable torque spectrum associated with OBC as
\begin{equation}
S^{\rm obc}_{\tau} = S_{\tau}^{\rm SQL} + S_{\tau}^{\rm OM} \ge S_{\tau}^{\rm SQL},
\label{Staubac}
\end{equation}
where $S_{\tau}^{\rm OM} = 4 \hbar \Omega_{\rm m} \Gamma_{\rm OM} I (\bar{n}_{\rm min} +1)$ is the contribution to the minimum resolvable torque spectrum resulting from optomechanical back-action, physically manifest as fluctuations in the radiation pressure force of the cooling laser due to photon shot noise. Therefore, OBC has the counterintuitive effect of increasing the minimum resolvable torque spectrum, as any reduction in phonon occupancy of the resonator is negated by an accompanying increase in the mechanical damping rate.  Thus, a torsional optomechanical resonator must be passively cooled towards ground state occupation in order to reach its standard quantum limit of torque sensitivity.

To meet this challenge, we have designed the optomechanically-detected torsional nanomechanical resonator shown in Fig.~\ref{fig1}, operating inside a dilution refrigerator at bath temperatures down to 17 mK.  Details of the cryogenic optomechanical system can be found elsewhere \cite{MacDonald2015}.  One merit of this system is that the use of a dimpled-tapered optical fiber \cite{Hauer2014} results in a full system optical detection efficiency of $\eta = 32\%$, which aids in the low-power optical measurements described below.

\begin{figure*}[b]
%%%%%%%%%%%%%%%%%   F I G U R E  3   %%%%%%%%%%%%%%%%%%
\centerline{\includegraphics[width=6.0in]{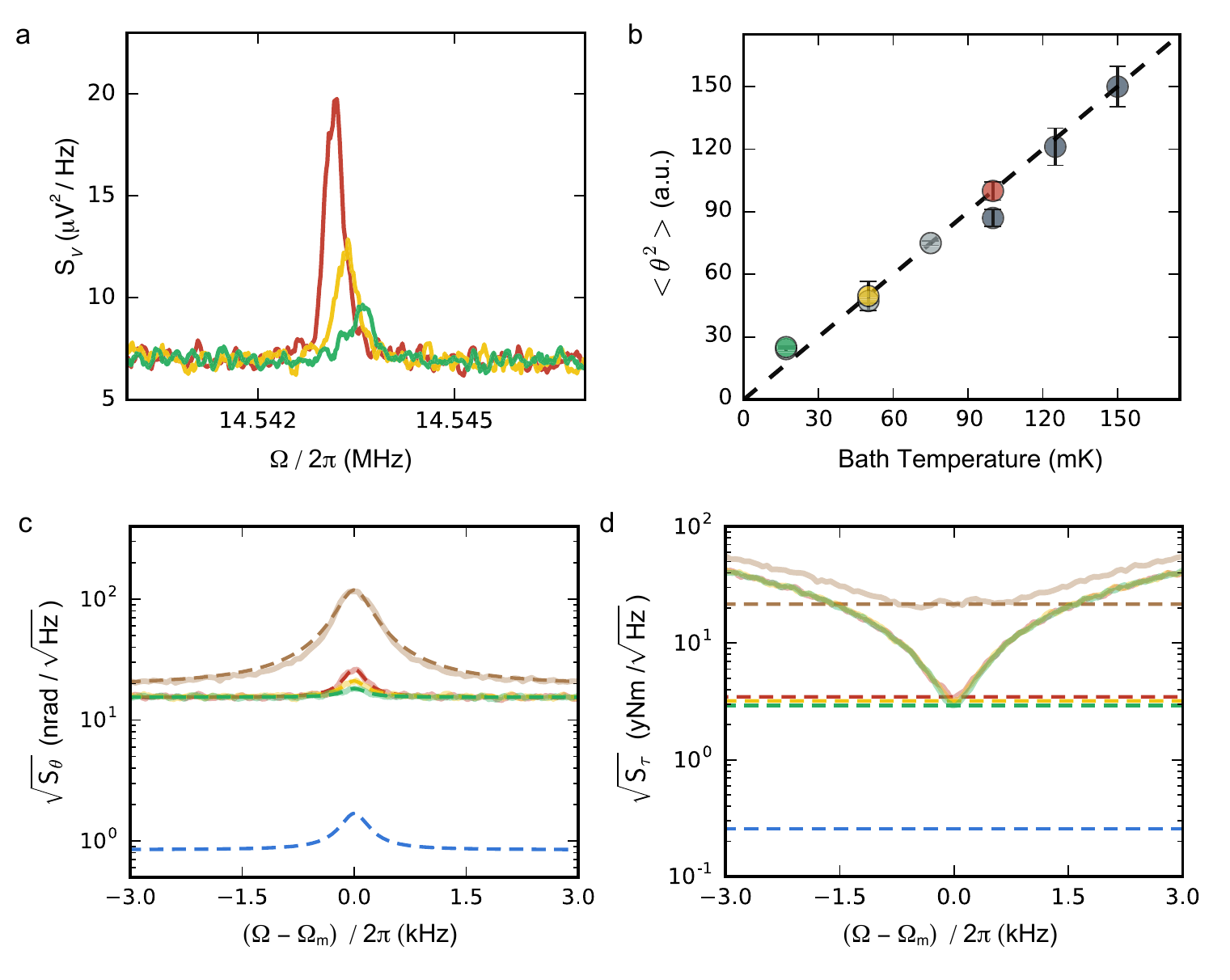}}
%%%%%%%%%%%%%%%%%%%%%%%%%%%%%%%%%%%%%%%%%%%%%
\caption{{\label{fig3}}{\bf Data at mK temperatures.} \textbf{a}, Power spectral densities of the torsional resonance at 100 mK (red), 50 mK (yellow), and 17 mK (green) dilution refrigerator temperatures. To minimize optical heating, we use low input power at the device (1.26 $\mu$W, corresponding to approximately 200 intracavity photons) and a low duty cycle measurement (see Supplementary Information). \textbf{b}, The integrated area under the power spectral density for three data runs, normalized to the highest temperature in each run. Slow system drift limits data acquisition to three temperatures per run.  Colored data points correspond to the measurements in (a), while the light and dark grey points are  additional runs.  Linearity between the integrated areas and the bath temperature implies adequate mode thermalization down to 25 mK, corresponding to an average phonon occupancy of $\braket{n} =35$. The dashed line has slope unity, representing a well-thermalized mode. \textbf{c}, Calibrated angular displacement spectral densities of the torsional mode at cryogenic temperatures (4.2 K brown, 100 mK red, 50 mK yellow, 25 mK green) with fits to equation \eqref{Stot} as dashed lines, along with the spectrum corresponding to zero-point fluctuations at the SQL calculated from device parameters in blue.  The measurement imprecision is higher than at the SQL due to the low optical power used to limit heating. \textbf{d}, Calibrated torque sensitivities corresponding to the measurements in (a,c). Dashed lines are the resonant torque sensitivity, reaching 2.9 yNm/$\sqrt{\rm Hz}$ at $T = 25$ mK: just over a factor of ten above its quantum-limited value of 0.26 yNm/$\sqrt{\rm Hz}$. {\label{fig3}} }
\end{figure*}

The torsional mechanical mode, at $\Omega_{\rm m}/2\pi = 14.5$ MHz, has low effective mass \cite{Hauer2013}, $m = 123$ fg (geometric mass of 1.14 pg); low effective moment of inertia, $I = 774$ fg$\cdot \mu$m$^2$; and low mechanical dissipation, $\Gamma/2\pi = 340$ Hz.  To sensitively measure the small angular motion of the device, we engineer a large angular (linear) dispersive optomechanical coupling, $G_\theta = {d\omega_{\rm c}}/{d\theta}$ ($G_x = {d\omega_{\rm c}}/{dx}$ - see Supplementary Information). The arms of the torsional resonator arc along the optical disk, over one-sixth of its perimeter, resulting in $G_\theta = 3.4$ GHz/mrad ($G_x = 1.4$ GHz/nm), while separating the mechanical element from the bulk of the optical field as compared to optomechanical crystals \cite{Eichenfield2009,Riedinger2016,Meenehan2015}.  In addition, the optical resonance at $\omega_{\rm c} / 2 \pi$ = 187 THz, the modeshape of which is shown in Fig.~\ref{fig1}b, is over-coupled to the dimpled-tapered fiber, with $\kappa_\textrm{e}/2\pi=7.1 $ GHz and $\kappa_\textrm{i}/2\pi=2.6$ GHz, further reducing optical absorption in the mechanical element. Due to the small mass and low frequency of the torsional mode, its zero-point motion is relatively large with $\theta_{\rm zpf} = 27$ nrad ($x_\textrm{zpf} = 69$ fm), resulting in a single-phonon coupling rate of $g_0/2\pi = 15$ kHz and a single photon cooperativity of $C_0 = 3\times10^{-4}$. 

\begin{figure*}[t]
%%%%%%%%%%%%%%%%%   F I G U R E  4   %%%%%%%%%%%%%%%%%%
\centerline{\includegraphics[width=6.0in]{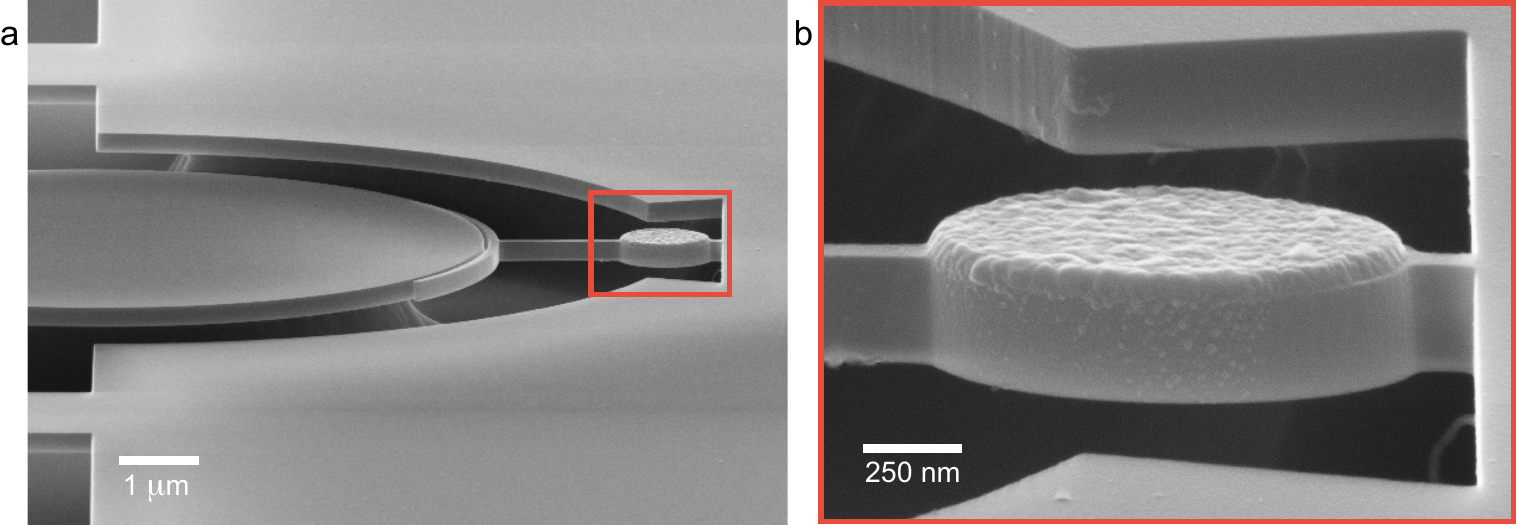}}
%%%%%%%%%%%%%%%%%%%%%%%%%%%%%%%%%%%%%%%%%%%%%
\caption{{\label{fig4}} {\bf Torsional optomechanical resonator with integrated micron-scale aluminum disk}. \textbf{a,b}, Fabrication demonstration of the optomechanical torque sensing platform via integration of a single mesoscopic aluminum disk (1.1 $\mu$m diameter, 45 nm thick).  An optomechanical torque sensitivity of 2.9 yNm/$\sqrt{\rm Hz}$ will not only enable measurements of single superconducting vortices \cite{Geim1997}, but also for the first time the mechanically detected dynamics \cite{Losby2015} of single vortices.  Room temperature measurements show that the optical and mechanical modes are unaffected by the presence of the metallic test sample.
{\label{fig4}} }
\end{figure*}

First, we find that using helium exchange gas to thermalize optomechanical resonators to 4.2 K is efficient, even at high optical input powers \cite{Wilson2014,MacDonald2016}, enabling measurement imprecision below the SQL as shown in Fig.~\ref{fig2}. On the other hand, when the exchange gas is removed for operation at millikelvin temperatures, light injected into the optomechanical resonator causes heating of the mechanical element \cite{Meenehan2015,Riedinger2016,MacDonald2016}.  To combat this parasitic optically-induced mechanical heating, we lower the duty cycle of the optomechanical measurement (see Supplementary Information).  Using a voltage-controlled variable optical attenuator, we apply a 20 ms optical pulse (limited by the mechanical linewidth) while continuously acquiring AC time-domain data, which are subsequently Fourier transformed to obtain mechanical spectra.  We then wait 120 s (corresponding to a duty cycle of 0.017\%) for the mechanical mode to re-thermalize to the bath.  To acquire sufficient signal-to-noise, this optical pulse sequence is repeated 100 times.  The resulting averaged power spectral densities, as shown in Fig.~\ref{fig3}a, are fit \cite{Hauer2013} to extract the area under the mechanical resonance, $\braket{\theta^2}$, which is proportional to the device temperature.  The mechanical mode temperature is confirmed by comparing the linearity of $\braket{\theta^2}$ with respect to the mixing chamber temperature, as measured by a fast ruthenium oxide thermometer referenced to a $^{60}$Co primary thermometer, Fig.~\ref{fig3}b.  We find that the mechanical mode is well thermalized to the bath, except near the fridge base temperature of 17 mK at which point the mechanical mode temperature is limited to 25 mK, corresponding to an average phonon occupancy of $\braket{n}=35$.

According to equation (4), $\braket{n}=35$ corresponds to a measured torque sensitivity a factor of six above the fundamental quantum torque sensitivity limit, when operating at the SQL.  Yet thermomechanical calibration of the displacement using the mechanical mode temperature reveals that the low optical powers used to limit heating result in measurement imprecision above the SQL, as seen in Fig.~\ref{fig3}c.  Specifically, we find that we have 90 quanta of added noise---instead of the ideal single quantum---dominated by measurement imprecision, placing the measured torque sensitivity eleven times above its standard quantum limit of 0.26 yNm/$\sqrt{\rm Hz}$.  Nonetheless, calibration in terms of torque sensitivity, as seen in Fig.~\ref{fig3}d, reveals that at 25 mK we reach 2.9 yNm/$\sqrt{\textrm{Hz}}$, a 270-fold improvement over previous generations of optomechanical torque sensors \cite{Kim2013,Wu2014} and more than a 130-fold improvement from the room temperature sensitivity of the same device.  

To give some idea of a what a torque sensitivity of 2.9 yNm/$\sqrt{\textrm{Hz}}$ enables, imagine the experimental scenario described in Refs.~\cite{Davis2010,Losby2015}, where the torsional mode is driven on resonance by a torque generated from an AC magnetic field, $\textbf{H}$, orthogonal to both the magnetic moment of the test sample, $\bf{\mu}$, and the torsion axis, leading to $\tau=\mu \times \textbf{H}$.  A reasonable drive field of 1 kA/m would imply the ability to resolve approximately 230 single electron spins within a 1 Hz measurement bandwidth.  Excitingly, this sensitivity unleashes the advanced toolbox of mechanical torque spectroscopy \cite{Davis2010,Burgess2013,Losby2015} on mesoscopic superconductors, previously limited to static measurements of bulk magnetization via ballistic Hall bars with $\sim$10$^3$ electron spin sensitivity \cite{Geim1997}.  As a first step in this direction, we show in Fig.~\ref{fig4} a prototype optomechanical torsional resonator with a single micron-scale aluminum disk integrated onto its landing pad, fabricated via secondary post-release e-beam lithography \cite{Diao2013}.  Room-temperature optomechanical measurements of devices with integrated aluminum disks in this geometry reveal that neither the optics or mechanics are degraded by the presence of the metal.  Future integration of bias and drive fields into our cryogenic system will enable measurements of the dynamical modes \cite{Losby2015} of single superconducting vortices \cite{Geim1997}, and even the paramagnetic resonance of electrons trapped within the silicon device itself.

\section{Methods}

The device was fabricated from silicon-on-insulator (SOI) (silicon thickness of 250 nm on a 3 $\mu$m buried oxide) using ZEP-520a e-beam resist on a RAITH-TWO 30 kV system, followed by an SF$_6$ reactive-ion etch to transfer the pattern to the SOI. The device was released from the oxide by 26 minutes of BOE wet-etch followed by a dilute HF dip for 1 minute and subsequent critical-point drying.

The prototype shown in Fig.~\ref{fig4} was fabricated to test the optical and mechanical effects of an Al film on the landing-pad. Here the same nanofabrication procedure is followed, including the BOE wet-etch, yet instead of drying the chip it was transferred to an acetone bath. The procedure for post-alignment then follows Ref.~\cite{Diao2013}: PMMA A8 resist was added to the acetone droplet on the chip to replace the acetone with resist. After spinning the resist, write-field alignments were done using pre-patterned alignment marks. High-purity aluminum (99.9999 \%) was deposited using an e-gun evaporation system at 2 $\times$ $10^{-7}$ Torr. After depositing 45 nm of Al, lift-off was performed by soaking in an NMP (\textit{N}-methyl-2-pyrrolidone) solvent for 1 hour at 80 $^{\circ}$C. Afterwards the chip was transferred to IPA and then critical-point dried.   

Optomechanical measurements in the dilution refrigerator are performed as in Refs.\,\cite{MacDonald2015,MacDonald2016}, yet using a dimpled tapered fiber \cite{Hauer2014} with a $\sim$50 $\mu$m diameter dimple, and the pulse scheme described in the Supplementary Information.

\section{Acknowledgements}
This work was supported by the University of Alberta, Faculty of Science; Alberta Innovates Technology Futures (Tier 3 Strategic Chair); the Natural Sciences and Engineering Research Council, Canada (RGPIN 401918 \& EQPEG 458523); the Canada Foundation for Innovation; and the Alfred P. Sloan Foundation. B.D.H. acknowledges support from the Killam Trusts.

\section{Author contributions}
P.H.K., B.D.H., and J.P.D. conceived and designed the experiment.  P.H.K. performed all nanofabrication.  P.H.K., B.D.H., and F.S. constructed and operated the cryogenic system.  C.D. wrote all data acquisition software.  P.H.K., B.D.H., and C.D. acquired and analyzed the data.  B.D.H. performed torque sensitivity calculations.  P.H.K., B.D.H., and J.P.D. wrote the manuscript.  All authors contributed to manuscript and figure editing.

%\section{Competing financial interests}  The authors declare no competing financial interests.

\section{Supplementary Information for Approaching the Standard Quantum Limit of Mechanical Torque Sensing}

\setcounter{equation}{0}

\renewcommand{\thefigure}{S\arabic{figure}}
\renewcommand{\theequation}{S\arabic{equation}}
\renewcommand{\thetable}{S\arabic{table}}
\renewcommand*{\citenumfont}[1]{S#1}
\renewcommand*{\bibnumfmt}[1]{[S#1]}

\subsection{Definitions: Fourier Transforms and Single Sided Spectral Densities}
\label{Def}

We define the Fourier transform of a time-dependent quantity $A(t)$, from which we obtain its spectral representation $A(\Omega)$, as
\begin{equation}
A(\Omega) = \int^{\infty}_{-\infty} A(t) e^{i \Omega t} dt,
\label{Fourdef}
\end{equation}
with the inverse Fourier transform being
\begin{equation}
A(t) = \frac{1}{2 \pi} \int^{\infty}_{-\infty} A(\Omega) e^{-i \Omega t} d\Omega.
\label{invFourdef}
\end{equation}
The double-sided spectral density of $A(t)$, $S_{AA}(\Omega)$, is then defined as the Fourier transform of its autocorrelation function $R_{AA}(t) = \braket{A(t)A(0)}$ \cite{hauer,clerk}
\begin{equation}
S_{AA}(\Omega) = \int^{\infty}_{-\infty} R_{AA}(t) e^{i \Omega t} dt.
\label{PSD2def}
\end{equation}
This function specifies the intensity of the signal $A(t)$ at a given frequency and is defined for all frequencies, both positive and negative, with the total energy of $A(t)$ being obtained by integrating over this entire spectral domain.

From this double-sided spectral density function, we can also introduce the symmetrized single-sided spectral density \cite{clerk,wilson}, defined strictly for positive frequencies as
\begin{equation}
S_A(\Omega) = S_{AA}(\Omega) + S_{AA}(-\Omega).
\label{PSD1def}
\end{equation}
Note that if $S_{AA}(\Omega)$ is an even function with respect to frequency, then equation \eqref{PSD1def} simply becomes $S_A(\Omega) = 2 S_{AA}(\Omega)$. It is this single-sided spectral density that is most often associated with experimentally measured spectra and will therefore be the spectral density we choose to use here.

\subsection{Torsional Mechanics}
\label{Tors}

Generally, three-dimensional motion of a mechanical resonator undergoing simple harmonic motion can be described by the displacement function
\begin{equation}
\mathbf{u}(\mathbf{r},t) = x(t) \mathbf{q}(\mathbf{r}),
\label{ux}
\end{equation}
\noindent where $x(t)$ is the time-dependent amplitude of motion and $\mathbf{q}(\mathbf{r})$ describes the spatially varying modeshape of the extended resonator structure \cite{hauer}. Here, we choose to normalize $\mathbf{q}(\mathbf{r})$ such that it is unitless and has a value of unity at its maximum ({\it i.e.}~max$| \mathbf{q}(\mathbf{r})|$ = 1).  In this way, $x(t)$ parametrizes the device's maximum amplitude of motion in units of displacement. A simple, yet effective, model for the dynamics of the system is that of a damped harmonic oscillator, whereby $x(t)$ obeys the equation of motion 
\begin{equation}
\ddot{x}(t) + \Gamma \dot{x}(t) + \Omega_{\rm m}^2 x(t) = \frac{f(t)}{m},
\label{xeom}
\end{equation}
\noindent where $\Gamma$, $\Omega_{\rm m}$, and $m$ are the mechanical resonator's damping rate, resonant angular frequency, and effective mass, respectively, with $f(t)$ being the external driving force of the system \cite{hauer}.

\begin{figure*}[t]
%%%%%%%%%%%%%%%%%  S I   F I G U R E  1   %%%%%%%%%%%%%%%%%%
\centerline{\includegraphics[width=\columnwidth]{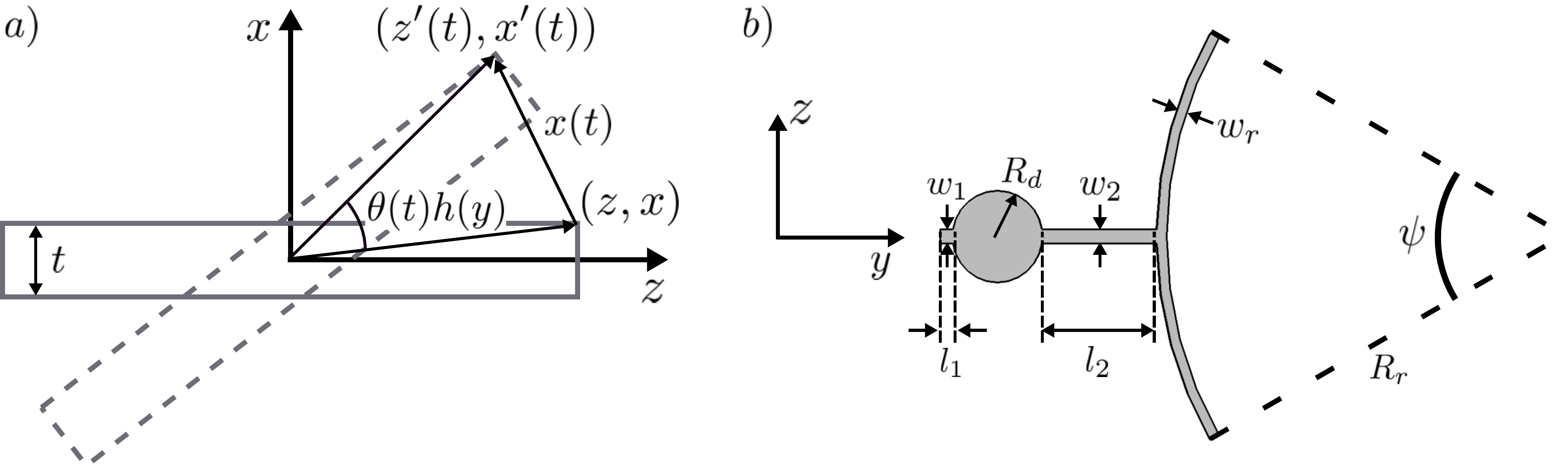}}
%%%%%%%%%%%%%%%%%%%%%%%%%%%%%%%%%%%%%%%%%%%%%
\caption{{\bf Out-of-plane torsional motion and resonator dimensions}.  \linespread{1.0} {\bf a}, A schematic depicting the out-of-plane displacement of a simple torsional mode. {\bf b}, A top-down view of the torsional resonator geometry used in this work with the critical dimensions labeled. The measured numerical value for each dimension can be found in Table \ref{Tab1}.
{\label{SIFig1}} }
\end{figure*}

For the specific case of torsional mechanics, it is more natural to instead characterize the resonator's motion in terms of an angular displacement, $\theta(t)$, from a pre-determined rotation axis, chosen here to be the $y$-axis. In this case, the displacement of the resonator will be confined to the $zx$-plane as defined by $\theta(t) h(y)$ (see Fig.~\ref{SIFig1}a), where we have introduced the scaling function $h(y) \in [0,1]$ that simply determines the magnitude of $\theta(t)$ along the $y$-axis. Thus, we have assumed the simplified, yet effective, model of torsional mechanics whereby there is no motion in the $y$-direction \cite{timoshenko,sokolnikoff}. For rigid, linear rotation, the time-dependent displacements $z'(t)$ and $x'(t)$ from an equilibrium point $(z,x)$ will then be given by
\begin{equation}
\begin{split}
\Delta z(t) &= z'(t) - z = z \left[ \cos(\theta(t) h(y)) - 1 \right] - x \sin(\theta(t) h(y)) \approx -\theta(t) h(y) x, \\
\Delta x(t) &= x'(t) - x = z \sin(\theta(t) h(y)) + x \left[ \cos(\theta(t) h(y)) - 1 \right] \approx \theta(t) h(y) z,
\label{dispxz}
\end{split}
\end{equation}
where we have used the small angle approximation $\theta(t) \ll 1$, valid for nanomechanical torsional resonators \cite{kim,wu}. Using the relations in equation \eqref{dispxz}, the displacement function in equation \eqref{ux} takes on the new form
\begin{equation}
\mathbf{u}(\mathbf{r},t) = \theta(t) h(y) z \hat{x} - \theta(t) h(y) x \hat{z} = \theta(t) \mathbf{p}(\mathbf{r}).
\label{uth}
\end{equation}
Here we have introduced a new modeshape function $\mathbf{p}(\mathbf{r}) = h(y) \left(z \hat{x} - x \hat{z} \right)$, which carries units of displacement.

We can now obtain a relationship between $\theta(t)$ and $a(t)$, by equating ${\rm max} |\mathbf{u}(\mathbf{r},t)|$ for both the linear and angular motion, given by equations \eqref{ux} and \eqref{uth} respectively, resulting in
\begin{equation}
x(t) = r_{\rm max} \theta(t),
\label{xtoth}
\end{equation}
where we have used the fact that ${\rm max}|\mathbf{p}(\mathbf{r})| = r_{\rm max} = \sqrt{x_{\rm max}^2 + z_{\rm max}^2}$, with $x_{\rm max}$ ($z_{\rm max}$) being the maximum extent of the resonator in the $x$ ($z$) direction. From this relation between linear and angular displacement, we can derive an equation of motion for $\theta(t)$ from equation \eqref{xeom} as
\begin{equation}
\ddot{\theta}(t) + \Gamma \dot{\theta}(t) + \Omega_{\rm m}^2 \theta(t) = \frac{\tau(t)}{I},
\label{theom}
\end{equation}
where we have now introduced the resonator's effective moment of inertia $I = r_{\rm max}^2 m$ and the external time-dependent torque $\tau(t) = r_{\rm max} f(t)$.

\begin{table}[b!]
\begin{tabular}{ ccc }
\hline
Measured Parameters & Material Parameters (Si) & Calculated Quantities \\  
\hline
\hline
$l_1$ = 155 nm & $\rho = 2329$ kg$\cdot$m$^3$ & $\kappa_1$ = 1.09 $\times$ 10$^{-10}$ N$\cdot$m \\
$w_1$ = 175 nm & $E = 170$ GPa & $\kappa_2$ = 1.16 $\times$ 10$^{-11}$ N$\cdot$m \\
$l_2$ = 1.46 $\mu$m & $\nu = 0.28$ & $I_{\kappa_1}$ = 1.23 $\times$ 10$^{-31}$ kg$\cdot$m$^2$ \\
$w_2$ = 175 nm & & $I_{\kappa_2}$ = 1.15 $\times$ 10$^{-30}$ kg$\cdot$m$^2$ \\
$w_r$ = 125 nm & & $I_1$ = 5.14 $\times$ 10$^{-29}$ kg$\cdot$m$^2$ \\
$t$ = 250 nm & & $I_2$ = 8.06 $\times$ 10$^{-28}$ kg$\cdot$m$^2$ \\
$R_d$ = 570 nm & & ~ \\
$R_r$ = 4.88 $\mu$m & & ~ \\
$\psi$ = 60.2 deg & & ~ \\
$r_{\rm max}$ = 2.51 $\mu$m & & ~ \\
\hline
\end{tabular}
\caption{Measured and calculated parameters for the torsional device studied in this work.}
\label{Tab1}
\end{table}

As it is often the case that measurements of mechanical motion are performed in the frequency domain, it is fruitful to Fourier transform equation \eqref{theom} to obtain a spectral representation of the device's angular motion as
\begin{equation}
\theta(\Omega) = \chi(\Omega) \tau(\Omega),
\label{xeomFour}
\end{equation}
\noindent where $\theta(\Omega)$ and $\tau(\Omega)$ are the Fourier transforms of $\theta(t)$ and $\tau(t)$. We have also introduced the generalized angular displacement susceptibility, $\chi(\Omega)$, which relates the angular displacement to the driving torque in frequency space and is given by
\begin{equation}
\chi(\Omega) = \frac{1}{I \left( \Omega_{\rm m}^2 - \Omega^2 - i \Omega \Gamma \right)}.
\label{chi}
\end{equation}
\noindent Furthermore, we can use this expression to relate the single-sided angular displacement spectral density $S_\theta(\Omega)$ to the single-sided spectral density of the driving torque as 
\begin{equation}
S_\theta (\Omega) = |\chi(\Omega)|^2 S_\tau(\Omega).
\label{PSDeq}
\end{equation}
Therefore, if we are able to measure the angular displacement spectrum of a torsional resonator, the torque acting on the system can be inferred via the system's angular displacement susceptibility.

\subsection{Effective Moment of Inertia}
\label{Effmom}

As we shall see, the effective moment of inertia introduced in equation \eqref{theom} is the geometric parameter that sets the torque sensitivity of a given torsional resonator. An explicit expression for the effective moment of inertia, $I$, can be determined by investigating the potential energy of the torsional spring, in direct analogy to the method by which one calculates a mechanical mode's effective mass \cite{hauer}. For a torsional resonator with a position-dependent density $\rho(\mathbf{r})$, the potential energy of a single, infinitesimal element, with volume $dV$ and mass $dm = \rho(\mathbf{r}) dV$, will be given by
\begin{equation}
dU = \frac{1}{2} \Omega_{\rm m}^2 \theta^2(t)  |\mathbf{p}(\mathbf{r})|^2 dm.
\label{potelem}
\end{equation}
The total potential energy is then found to be
\begin{equation}
U = \frac{1}{2} I \Omega_{\rm m}^2 \theta^2(t),
\label{pottot}
\end{equation}
with the effective moment of inertia
\begin{equation}
I = \int |\mathbf{p}(\mathbf{r})|^2 dm = \int \rho(\mathbf{r}) h^2(y) \left( x^2 + z^2 \right) dV,
\label{effmom}
\end{equation}
where the integral performed over the entire volume of the device. We note that if $h(y) \approx 1$ over the extent of the resonator that contains the majority it's mass (physically corresponding to a large, wide torsion paddle), then we can see from equation \eqref{effmom} that $I \approx I_0$, where $I_0$ is the conventional, geometric moment of inertia of the device. In the following two subsections we will investigate this parameter, using both analytical methods and numerical simulation, for the resonator geometry discussed in this work.

\subsubsection{Analytical Model}
\label{Ana}

\begin{figure*}[t!]
%%%%%%%%%%%%%%%%%  S I   F I G U R E  2   %%%%%%%%%%%%%%%%%%
\centerline{\includegraphics[width=\columnwidth]{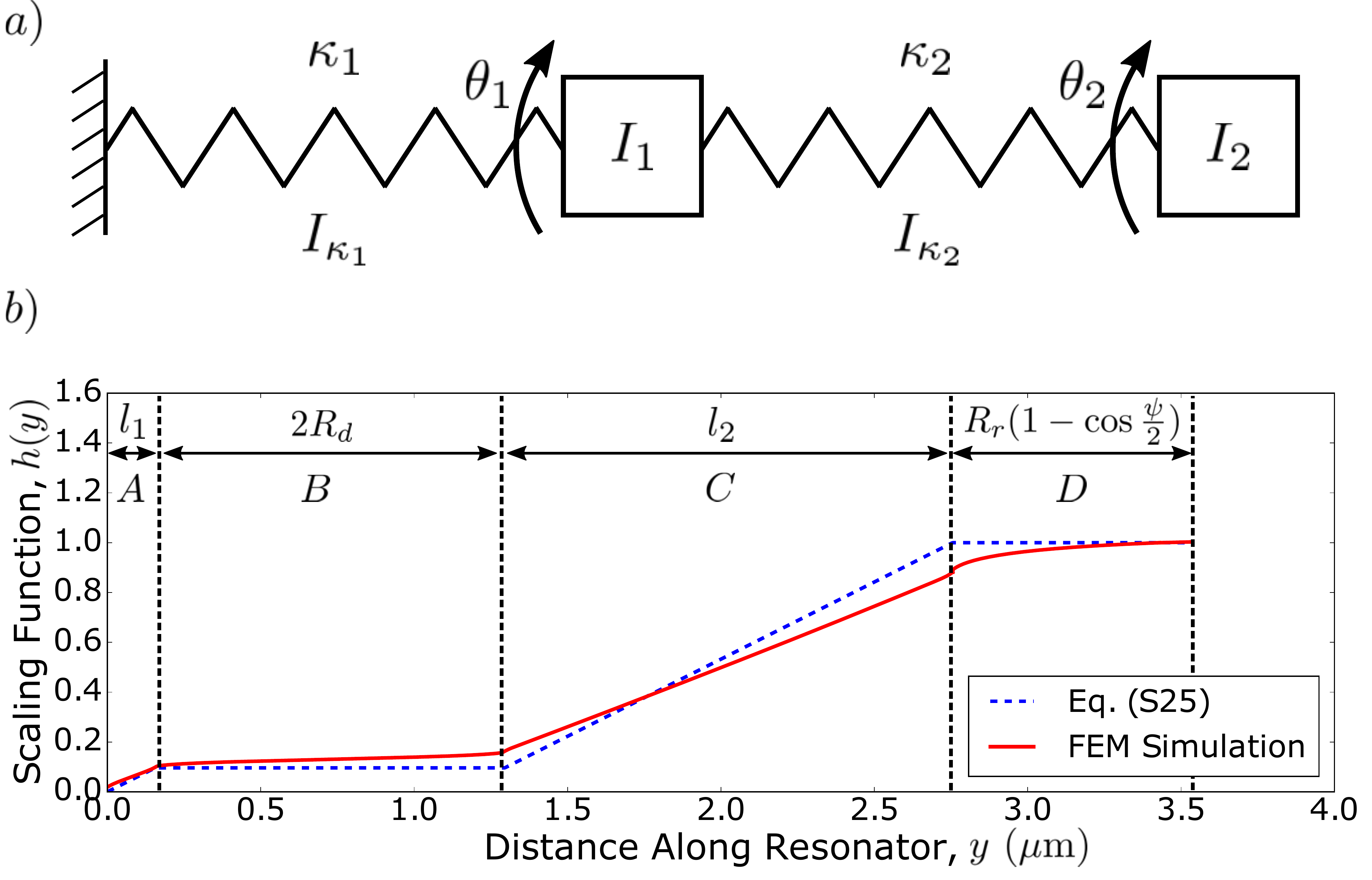}}
%%%%%%%%%%%%%%%%%%%%%%%%%%%%%%%%%%%%%%%%%%%%%
\caption{{\bf Coupled torsional model schematic and angular scaling function}. {\bf a}, A torsional ``mass-and-spring'' diagram illustrating the simple coupled oscillator model used in Sec.~\ref{Ana}. {\bf b}, Plot of the scaling function $h(y)$ using both the analytical model of equation \eqref{hysym} (blue - dashed) and FEM simulation (red - solid), with the following four regions of the resonator demarcated: $A$ - first torsion rod, $B$ - sample disk, $C$ - second torsion rod and $D$ - ring segment. 
{\label{SIFig2}} }
\end{figure*}

From equation \eqref{effmom}, we see that in order to calculate the effective moment of inertia of a torsional mode, we need to determine it's mechanical modeshape, $\mathbf{p}(\mathbf{r})$, or more specifically, the scaling function, $h(y)$. We begin by developing a simple analytical model for the resonator shown in Fig.~\ref{SIFig1}b. For this geometry, we treat the system as two coupled, torsional resonators (see Fig.~\ref{SIFig2}a), one punctuated by the sample disk, the other by the ring segment used to couple to the optical disk, each with it's own torsional spring constant, $\kappa_i$, and (geometric) moment of inertia, $I_i$. For the device considered here, the torsion rods have a simple rectangular cross-section, such that the torsional spring constants will be given by
\begin{equation}
\kappa_i = \frac{\beta t w_i^3 E}{2 l_i \left(1 + \nu \right)},
\label{kappa}
\end{equation}
where $l_i$ and $w_i$ are the length and width of the torsion rod, and $t$, $E$ and $\nu$ are the thickness, Young's modulus and Poisson's of the device, while $\beta$ is a numerical coefficient given by
\begin{equation}
\beta = \frac{1}{3} \left[ 1 - \frac{192}{\pi^5} \frac{w_i}{t} \sum^{\infty}_{n} \frac{1}{n^5} \tanh \left( \frac{n \pi t}{2 w_i} \right) \right],
\label{beta}
\end{equation}
where $n$ are positive odd integers \cite{bao,timoshenko}. Note that we have assumed $t > w_i$, as this is the case for the device studied here.

The moment of inertia of the sample disk is found from its geometry to be
\begin{equation}
I_1 = \frac{m_d}{4} \left( R_d^2 + \frac{t^2}{3} \right),
\label{Idisk}
\end{equation}
where $R_d$ and $m_d = \rho \pi R_d^2 t$ are the radius and mass of the sample disk, with $\rho$ being the density of the device. On the other hand, the moment of inertia of the ring segment is given by
\begin{equation}
I_2 = \frac{\rho t w_r \left( 2R_r + wr \right)}{8} \left[ \left( \left(R_r + w_r \right)^2 + R_r^2 \right) \left( \psi - \sin \psi \right) + \frac{\psi t^2}{3} \right],
\label{Iring}
\end{equation}
where $w_r$, $R_r$ and $\psi$ are the width, radius of curvature and sector angle of the ring segment as shown in Fig.~\ref{SIFig1}b.

With the above torsional spring constants and moments of inertia, the equations of motion for the device (excluding damping for simplicity) are then given by
\begin{equation}
\begin{split}
I_1 \ddot{\theta}_1 &= -\kappa_1 \theta_1 - \kappa_2 \left( \theta_1 - \theta_2 \right), \\
I_2 \ddot{\theta}_2 &= -\kappa_2 (\theta_2 - \theta_1).
\label{coupEOM}
\end{split}
\end{equation}
where $\theta_1$ ($\theta_2$) is the angular displacement of the sample disk (ring segment), as demonstrated schematically in Fig.~\ref{SIFig2}a. Fourier-transforming these equations of motion, such that $\ddot{\theta}_i = - \Omega^2 \theta_i$, we can rewrite them in matrix form as $A \Theta = 0$, with $A$ and $\Theta$ given by
\begin{equation}
A=
 \left[ \begin{array}{cc}
\Omega^2 - \frac{\kappa_1 + \kappa_2}{I_1} & \frac{\kappa_2}{I_1} \\
\frac{\kappa_2}{I_2} & \Omega^2 - \frac{\kappa_2}{I_2}
\end{array} \right],~~~\Theta = \left[ \begin{array}{cc}
\theta_1 \\
\theta_2
\end{array} \right].
\label{Amat}
\end{equation}
Solving this system of equations, we find the eigenfrequencies of this coupled system to be
\begin{equation}
\Omega_\pm = \sqrt{\frac{1}{2} \left( \frac{\kappa_2}{I_2} + \frac{\kappa_1 + \kappa_2}{I_1} \right) \pm \frac{1}{2} \sqrt{\left( \frac{\kappa_2}{I_2} + \frac{\kappa_1 + \kappa_2}{I_1} \right)^2 - \frac{4\kappa_1 \kappa_2}{I_1 I_2}}},
\label{freqcoup}
\end{equation}
where $\Omega_-$ ($\Omega_+$) corresponds to the symmetric (antisymmetric) torsional mode. Here we focus on the symmetric mode (as this is the mode examined in this work), which for the experimental parameters given in Table~\ref{Tab1} has a predicted frequency of $\Omega_-/2 \pi = 18.1$ MHz, somewhat larger than the experimentally measured mechanical resonance frequency of $\Omega_{\rm m} / 2 \pi = 14.5$ MHz.

Inserting the analytical expression for the eigenfrequency of the symmetric mode into the system of equations given by equation \eqref{Amat}, we obtain $\theta_2$ in terms of $\theta_1$ as
\begin{equation}
\theta_2 = \frac{1}{2} \left( - \frac{I_1}{I_2} + \frac{\kappa_1 + \kappa_2}{\kappa_2} + \sqrt{ \left( \frac{I_1}{I_2} + \frac{\kappa_1 + \kappa_2}{\kappa_2} \right)^2 - \frac{4 \kappa_1 I_1}{\kappa_2 I_2}} \right) \theta_1 \approx \frac{\kappa_1 + \kappa_2}{\kappa_2} \theta_1,
\label{relth}
\end{equation}
where we have made the experimentally relevant approximation $I_1 \ll I_2$.

We can now use the relative angular displacements of the sample disk and ring segments given in equation \eqref{relth} to determine the modeshape scaling function $h(y)$. To do this, we assume the simplest imaginable torsional modeshape \cite{timoshenko,sokolnikoff}, where the mechanical device is rigidly clamped at one end ($\theta = 0$), with the angle of deflection increasing linearly along the torsion rods, while remaining constant over both the sample disk and ring segment. In this case, $h(y)$ is given by the piecewise function
\begin{equation}
h(y) = \left\{
     \begin{array}{lr}
       \frac{\kappa_2}{\kappa_1 + \kappa_2} \frac{y}{l_1}, & y \in A~ \\
       \frac{\kappa_2}{\kappa_1 + \kappa_2}, & y \in B~ \\
       \frac{\kappa_1}{\kappa_1 + \kappa_2} \frac{y - l_1 - 2 R_d}{l_2} + \frac{\kappa_2}{\kappa_1 + \kappa_2}, & y \in C~ \\
       1, & y \in D.
     \end{array}
   \right. 
\label{hysym}
\end{equation}
A plot of this function using the measured/calculated device parameters (as given in Table \ref{Tab1}) can be seen in Fig.~\ref{SIFig2}b, where we have also defined the regions $A$-$D$.

Finally, by inputting equation \eqref{hysym} into equation \eqref{effmom}, we can determine the effective moment of inertia for this simple analytical model to be
\begin{equation}
\begin{split}
I &= \left( \frac{\theta_1}{\theta_2} \right)^2 \frac{I_{\kappa_1}}{3} + \left[ 1 + \frac{\theta_1}{\theta_2} + \left( \frac{\theta_1}{\theta_2} \right)^2 \right] \frac{I_{\kappa_2}}{3} + \left( \frac{\theta_1}{\theta_2} \right)^2 I_1 + I_2 \\
&\approx \left( \frac{\kappa_2}{\kappa_1 + \kappa_2} \right)^2 \frac{I_{\kappa_1}}{3} + \left[ 1 + \frac{\kappa_2}{\kappa_1 + \kappa_2} + \left( \frac{\kappa_2}{\kappa_1 + \kappa_2} \right)^2 \right] \frac{I_{\kappa_2}}{3} + \left( \frac{\kappa_2}{\kappa_1 + \kappa_2} \right)^2 I_1 + I_2 \approx I_2,
\label{effmomana}
\end{split}
\end{equation}
where $I_{\kappa_1}$ and $I_{\kappa_2}$ are the geometric moments of inertia corresponding to the two torsion rods with spring constants $\kappa_1$ and $\kappa_2$ ($I_{\kappa_i} = \rho l_i w_i t (w_i^2 + t^2)/12$), and we have again used the experimentally valid approximation that $I_2 \gg I_1,I_{\kappa_1},I_{\kappa_2}$. For the device studied here, the analytical model predicts an effective moment of inertia $I = 807$ fg$\cdot \mu$m$^2$.

\subsubsection{Finite Element Method Simulation}
\label{FEM}

While the analytical model of the previous section allows for a qualitative understanding of the torsional modeshape of our resonator, a much more accurate modeshape, and therefore effective moment of inertia, can be determine using finite element method (FEM) simulations. An example of such a modeshape generated using COMSOL can be seen in the inset of Fig.~1d of the main text. From this simulated modeshape, the effective moment of inertia of the device can be calculated numerically using equation \eqref{effmom}, for which we find a value of $I = 774$ fg$\cdot \mu$m$^2$. It is this value that we use to calculate the torque sensitivities in the main text.

Furthermore, the scaling function $h(y)$ extracted from the FEM simulation is compared to its analytically determined counterpart given by equation \eqref{hysym} in Fig.~\ref{SIFig2}b, highlighting the deviation of the analytical model from the numerical one. This disparity is likely due to the fact that the analytical model ignores the resonator's support structure, as well as the elasticity of the material. As such, the analytical model overshoots the scaling function in the critical region of the ring segment, predicting a larger effective moment of inertia than the numerical model. 

\subsection{Limits on Continuous Linear Torque Measurements}
\label{Lim}

The limit on the sensitivity one can obtain by performing a continuous linear torque measurement using a mechanical resonator is set by the noise that will inevitably creep into the system, contaminating the measurement. Minimizing this noise will allow the system to resolve smaller torques, leading to an increase in sensitivity.

In order to determine the limiting torque noise, we consider the total angular displacement noise spectrum
\begin{equation}
S_\theta(\Omega) = S_\theta^{\rm qu}(\Omega) + S_\theta^{\rm imp}(\Omega) + S_\theta^{\rm ba}(\Omega),
\label{Stttot}
\end{equation}
which is a combination of the intrinsic noise due to the thermal and quantum fluctuations of the mechanical element, $S_\theta^{\rm qu}(\Omega)$, along with the imprecision, $S_\theta^{\rm imp}(\Omega)$, and back-action, $S_\theta^{\rm ba}(\Omega)$, noise spectra generated by the measurement apparatus.

The intrinsic angular noise spectrum can be determined using the (quantum) fluctuation-dissipation theorem \cite{braginsky}, expressed mathematically for the single-sided angular displacement as
\begin{equation}
S_\theta^{\rm qu}(\Omega) = 4 \hbar \left(\braket{n} + 1/2 \right) {\rm Im} \{ \chi(\Omega) \} = 4 \hbar \Omega \Gamma I \left(\braket{n} + 1/2 \right) | \chi(\Omega)|^2.
\label{QFDT}
\end{equation}
Comparing this expression to equation \eqref{PSDeq}, we can immediately identify the intrinsic quantum torque spectrum as $S_\tau^{\rm qu}(\Omega) = 4 \hbar \Omega \Gamma I \left(\braket{n} + 1/2 \right)$, where $\braket{n}$ is the phonon occupation of the torsional mode. For the case of thermal equilibrium with a bath at temperature $T$, the phonon occupation is given by the Bose-Einstein occupation factor $\braket{n} = \bar{n}_{\rm th} = (e^{\hbar \Omega/k_B T} - 1)^{-1}$. Note that we also have an addition of one-half to this thermal occupation corresponding to the ground state motion of the mechanical resonator. In the high-temperature limit, $\braket{n} \approx k_B T / \hbar \Omega \gg 1$, this ground state contribution can be neglected and $S_\tau^{\rm qu}(\Omega)$ reduces to the familiar classical white-noise torque spectrum of $S_\tau^{\rm cl} = 4 k_B T \Gamma I$ \cite{hauer}. 

Likewise, one can express the back-action angular noise spectrum as $S_\theta^{\rm ba}(\Omega) = |\chi(\Omega)|^2 S_\tau^{\rm ba}(\Omega)$, where we have now introduced a back-action torque noise spectrum, $S_\tau^{\rm ba}(\Omega)$. In general, this back-action torque spectrum, as well as the angular imprecision noise spectrum, $S_\theta^{\rm imp}(\Omega)$, will contain contributions from both classical technical noise and fundamental quantum noise, with the product of the two spectra obeying the Heisenberg uncertainty relation (for single-sided spectra) \cite{clerk,wilson,braginsky}
\begin{equation}
S_\tau^{\rm ba}(\Omega) S_\theta^{\rm imp}(\Omega) \ge \hbar^2.
\label{Heisineq}
\end{equation}
Note that equality in equation \eqref{Heisineq} corresponds to quantum-limited measurement noise ({\it i.e.} no classical noise).

We can now determine an equivalent torque noise spectrum from equation \eqref{Stttot} using the angular susceptibility of the system as
\begin{equation}
S_\tau(\Omega) = \frac{S_\theta(\Omega)}{|\chi(\Omega)|^2} = 4 \hbar \Omega \Gamma I \left( \braket{n} + n_{\rm imp}(\Omega) + n_{\rm ba}(\Omega) + 1/2 \right),
\label{Stautautot}
\end{equation}
It is this torque noise spectrum that sets the minimum resolvable torque, and hence the sensitivity, of our system. In equation \eqref{Stautautot}, we have introduced the equivalent noise quanta due measurement imprecision and back-action, $n_{\rm imp}(\Omega) = S_\theta^{\rm imp}(\Omega)/4 \hbar \Omega \Gamma I |\chi(\Omega)|^2$ and $n_{\rm ba}(\Omega) = S_\theta^{\rm ba}(\Omega)/4 \hbar \Omega \Gamma I |\chi(\Omega)|^2$, whose product can be found from equation \eqref{Heisineq} to obey
\begin{equation}
n_{\rm imp}(\Omega) n_{\rm ba}(\Omega) \ge \frac{1}{16 \Omega^2 \Gamma^2 I^2 |\chi(\Omega)|^2},
\label{Heisineqqaunta}
\end{equation}
with equality again corresponding to quantum-limited measurement noise. Note that at the mechanical resonance frequency, equation \eqref{Heisineqqaunta} reduces to the familiar form $n_{\rm imp}(\Omega_{\rm m}) n_{\rm ba}(\Omega_{\rm m}) \ge 1/16$ \cite{wilson}.

We now interest ourselves in determining the minimum possible torque noise spectrum as allowed by quantum mechanics. To do this, we first look to minimize the added measurement noise of our system
\begin{equation}
S_\theta^{\rm add}(\Omega) = S_\theta^{\rm imp}(\Omega) + S_\theta^{\rm ba}(\Omega) = S_\theta^{\rm imp}(\Omega) + |\chi(\Omega)|^2 S_\tau^{\rm ba}(\Omega).
\label{Stadd}
\end{equation}
By taking equality in the Heisenberg uncertainty relation of equation \eqref{Heisineq}, we find the optimal measurement noise spectra of $S_\theta^{\rm imp}(\Omega) = S_\theta^{\rm ba}(\Omega) = \hbar |\chi(\Omega)|$, or equivalently, $S_\tau^{\rm ba}(\Omega) = \hbar / |\chi(\Omega)|$, corresponding to the so-called standard quantum limit (SQL) of continuous position measurement \cite{clerk,braginsky}. Furthermore, by tuning to the mechanical resonance frequency, $|\chi(\Omega)|$ is maximized, thus minimizing the torque noise $S_\tau(\Omega)$ in equation \eqref{Stautautot} with respect to frequency. Returning to our effective quanta notation, we find that $n_{\rm imp}(\Omega_{\rm m}) = n_{\rm ba}(\Omega_{\rm m}) = 1/4$ at the SQL, such that the minimized torque noise is found to be
\begin{equation}
S_\tau^{\rm SQL} = S_\tau^0 (\braket{n} + 1),
\label{StauSQL}
\end{equation}
where $S_\tau^0 = 4 \hbar \Omega_{\rm m} \Gamma I$ is the fundamental torque noise spectrum associated with the continuous monitoring of a mechanical resonator in its quantum ground state at the SQL. It is this zero-point torque spectrum that sets the quantum limit for the minimum resolvable torque for a given device, half of which arises from the zero-point motion of the resonator, the other half from the quantum-limited measurement noise at the SQL, whereas the limit at finite temperature will be given by the quantity in equation \eqref{StauSQL}.

\subsection{Effect of Optomechanical Back-Action Cooling}
\label{BAC}

One method by which one might think to decrease the minimum resolvable torque spectrum as given by equation \eqref{StauSQL} is to reduce the phonon occupancy of the mechanical resonator using a cold damping method, such as optomechanical back-action cooling (OBC) \cite{wilson-rae,marquardt}. Unfortunately, as we will see below, due to the increase in the mechanical resonance's linewidth caused by such a process, the minimum resolvable torque spectrum in fact increases.

In OBC, photons trapped in an optical cavity impart a dynamical radiation pressure force on the mechanical resonator, increasing its intrinsic linewidth by an amount $\Gamma_{\rm OM}$ and reducing its average phonon occupancy according to
\begin{equation}
\braket{n} = \frac{\Gamma \bar{n}_{\rm th} + \Gamma_{\rm OM} \bar{n}_{\rm min}}{\Gamma + \Gamma_{\rm OM}},
\label{naveba}
\end{equation}
\noindent where $\bar{n}_{\rm min}$ is the minimum obtainable average phonon occupancy using this method \cite{wilson-rae,marquardt} ($\bar{n}_{\rm min} = \kappa / 4 \Omega_{\rm m}$ for $\kappa \gg \Omega_{\rm m}$, whereas $\bar{n}_{\rm min} = \kappa^2 / 16 \Omega_{\rm m}^2$ in the sideband resolved case of $\kappa \ll \Omega_{\rm m}$, with $\kappa$ being the linewidth of the optical cavity). Note that for $\Gamma_{\rm OM} = 0$ ({\it i.e.}~no optomechanical damping), $\braket{n} = \bar{n}_{\rm th}$ and thermal equilibrium is restored.

Inputting equation \eqref{naveba} into equation \eqref{StauSQL}, we obtain the minimum resolvable torque spectrum associated with OBC as
\begin{equation}
\begin{split}
S^{\rm obc}_\tau &= 4 \hbar \Omega_{\rm m} I (\Gamma + \Gamma_{\rm OM}) \left( \frac{\Gamma \bar{n}_{\rm th} + \Gamma_{\rm OM} \bar{n}_{\rm min}}{\Gamma + \Gamma_{\rm OM}} + 1 \right) \\ 
&= 4 \hbar \Omega_{\rm m} I (\Gamma \bar{n}_{\rm th} + \Gamma_{\rm OM} \bar{n}_{\rm min}) + 4 \hbar \Omega_{\rm m} I (\Gamma + \Gamma_{\rm OM}) \\
&= 4 \hbar \Omega_{\rm m} \Gamma I (\bar{n}_{\rm th} + 1) + 4 \hbar \Omega_{\rm m} \Gamma_{\rm OM} I (\bar{n}_{\rm min} +1) \\
&= S_\tau^{\rm SQL} + S_\tau^{\rm OM} \ge S_\tau^{\rm SQL},
\end{split}
\label{proofba}
\end{equation}
where $S_\tau^{\rm OM} = 4 \hbar \Omega_{\rm m} \Gamma_{\rm OM} I (\bar{n}_{\rm min} + 1)$ is the contribution to the minimum resolvable torque spectrum due to optomechanical back-action. Note that the inequality in the last line of equation \eqref{proofba} arises due to the fact that $S_\tau^{\rm OM} \ge 0$ for optomechanical damping (where $\Gamma_{\rm OM} \ge 0$), with equality corresponding to $\Gamma_{\rm OM} = 0 \Rightarrow S_\tau^{\rm OM} = 0$. Therefore, one can see that while the phonon occupancy of the mechanical mode decreases, there is in fact an increase in the minimum resolvable torque spectrum. Physically, this can be understood due to the fact that any apparent reduction in the minimum resolvable torque spectrum attributed to a decrease in the phonon occupancy of the resonator is nullified by the corresponding increase in the mechanical damping rate.

\subsection{Optomechanical Torque Transduction}
\label{OM}

In this work, torsional motion is transduced optomechanically, whereby the torsional motion of the mechanical resonator is coupled to photons confined in an optical cavity. Here, the coupling is dispersive in the sense that the resonance frequency of the optical cavity $\omega_{\rm c}(\theta(t))$ is a function of the angular motion $\theta(t)$, which can be expanded to first order as 
\begin{equation}
\omega_{\rm c}(\theta(t)) \approx \omega_c + G_\theta \theta(t),
\label{Gth}
\end{equation}
where $\omega_c$ is the unperturbed cavity resonance frequency and $G_\theta = d \omega_{\rm c} / d \theta$ is the angular optomechanical coupling coefficient. Note that we can use equation \eqref{xtoth} to relate $G_\theta$ to the standard optomechanical coupling coefficient for linear motion $G_x = d \omega_{\rm c} / dx$ as $G_\theta = r_{\rm max} G_x$. 

This mechanically induced optical frequency shift will result in amplitude and phase fluctuations in the optical signal transmitted through the cavity, both of which can be detected (using either a tuned-to-slope or homodyne measurement) as an AC voltage signal on a photodiode. This time-varying signal can then be converted into a voltage spectral density given by
\begin{equation}
S_V(\Omega) =  \alpha S_\theta(\Omega),
\label{PSDV}
\end{equation}
from which we can infer the angular displacement spectral density $S_\theta(\Omega)$ by determining the properly calibrated conversion coefficient $\alpha$, which has units of V$^2$/rad$^2$. In this work, this is done by thermomechanically calibrating the voltage spectrum using the method of Ref.~\cite{hauer}.

\begin{figure*}[t!]
%%%%%%%%%%%%%%%%%  S I   F I G U R E  3   %%%%%%%%%%%%%%%%%%
\centerline{\includegraphics[width=4.0in]{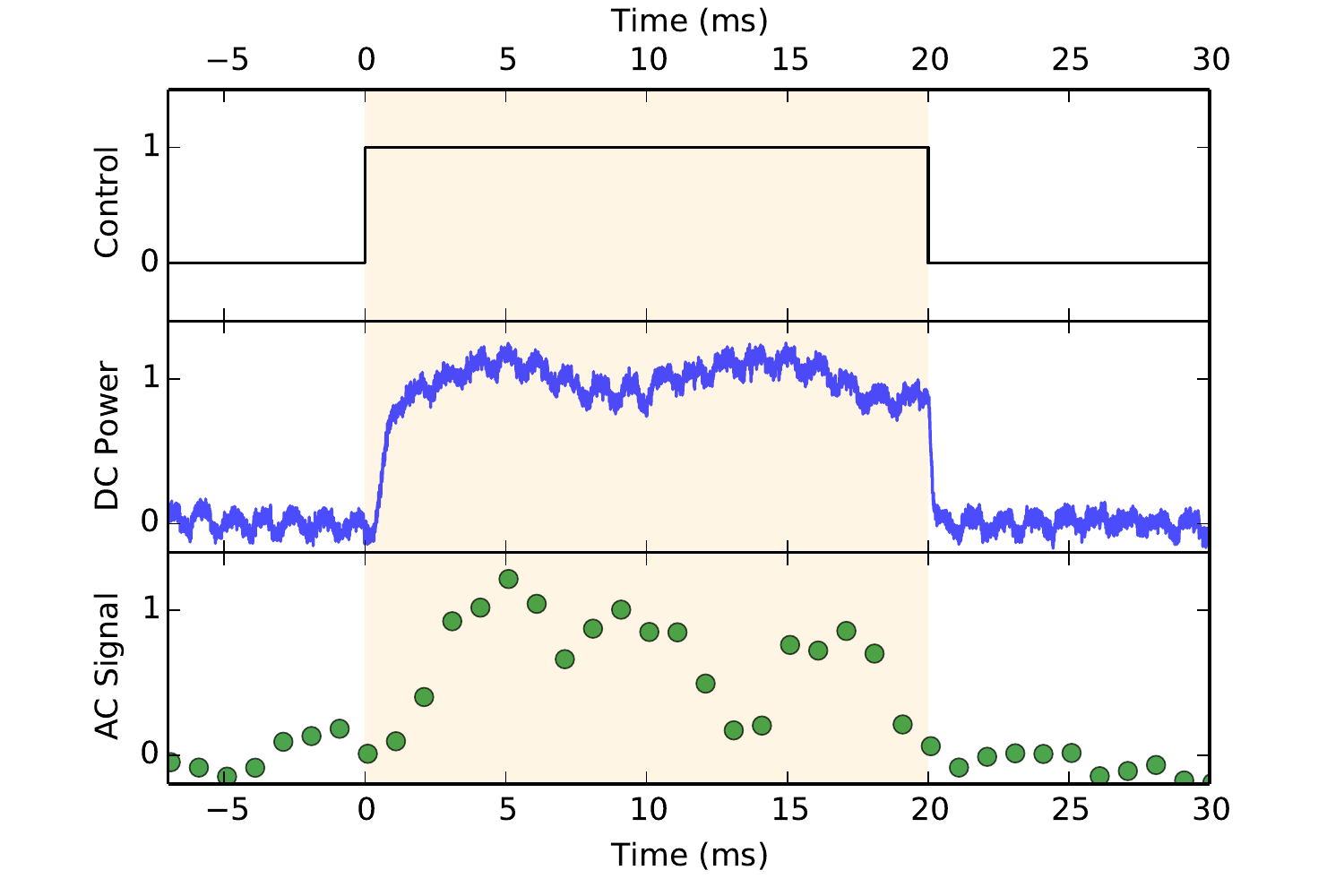}}
%%%%%%%%%%%%%%%%%%%%%%%%%%%%%%%%%%%%%%%%%%%%%
\caption{{\bf Low duty cycle measurements}. Light transmitted through the system is split into a DC component, which indicates the power in the optomechanical resonator, and an AC component.  The AC signal is bandpassed over a 1.4 kHz bandwidth centered at the mechanical resonance frequency.  The noise floor is estimated from 2 MHz of off-resonance signal and subtracted from the AC signal.  Here we show this AC signal sliding-averaged across 4.2 ms for every 1 ms.
{\label{SIFig3}} }
\end{figure*}

\subsection{Low Duty Cycle Measurements}
\label{Duty}
In the absence of helium exchange gas, the mechanical resonator is easily heated by injected optical power.  To circumvent this heating, low-duty cycle measurements are performed.  The optical power is adjusted using a voltage-controlled variable optical attenuator---labeled in Fig.~\ref{SIFig3} as control---and is pulsed on for 20 ms, as dictated by the mechanical damping rate at low temperatures, $\Gamma/2\pi = 340$ Hz.  We then wait 120 s without injecting any light, to allow for re-thermalization to the dilution refrigerator temperature. The AC spectrum from 100 of these pulse sequences is averaged to extract each of the data sets shown in Fig.~3 of the main text.

\end{document}